\documentclass[twocolumn,amssymb,nobibnotes,showpacs,superscriptaddress,aps,prx,longbibliography,floatfix]{revtex4-1}
\usepackage{graphicx,amsmath}

\usepackage[colorlinks,linkcolor=blue,urlcolor=blue,anchorcolor=blue,citecolor=blue]{hyperref}

\begin{document}
	
\title{Hybrid level anharmonicity and interference induced photon blockade in a two-qubit cavity QED system with dipole-dipole interaction}

\author{C. J. Zhu}
%\email[Corresponding author:]{cjzhu@tongji.edu.cn}
\affiliation{MOE Key Laboratory of Advanced Micro-Structured Materials, School of Physics Science and Engineering,Tongji University, Shanghai, China 200092}
\author{K. Hou}
\affiliation{MOE Key Laboratory of Advanced Micro-Structured Materials, School of Physics Science and Engineering,Tongji University, Shanghai, China 200092}
\affiliation{Department of Mathematics and Physics, Anhui JianZhu University, Hefei 230601,China}
\author{Y. P. Yang}
\email[Corresponding author:]{yang\_yaping@tongji.edu.cn}
\affiliation{MOE Key Laboratory of Advanced Micro-Structured Materials, School of Physics Science and Engineering,Tongji University, Shanghai, China 200092}
\author{L. Deng}
\email[Corresponding author:]{lu.deng@email.sdu.edu.cn}
\affiliation{MOE Key Laboratory of Advanced Micro-Structured Materials, School of Physics Science and Engineering,Tongji University, Shanghai, China 200092}
\affiliation{Center for Optics Research and Engineering (CORE), Shandong University, Qingdao, China 266237}
	
\begin{abstract}
We theoretically study a quantum destructive interference (QDI) induced photon blockade in a two-qubit driven cavity QED system with dipole-dipole interaction (DDI). In the absence of dipole-dipole interaction, we show that a QDI-induced photon blockade can be achieved only when the qubit resonance frequency is different from the cavity mode frequency. When DDI is introduced the condition for this photon blockade is strongly dependent upon the pump field frequency, and yet is insensitive to the qubit-cavity coupling strength. Using this tunability feature we show that the conventional energy-level-anharmonicity-induced photon blockade and this DDI-based QDI-induced photon blockade can be combined together, resulting in a hybrid system with substantially improved mean photon number and second order correlation function. Our proposal provides a non-conventional and experimentally feasible platform for generating single photons.
%\text{Keywords:}
\end{abstract}
	
\maketitle
	
\section{Introduction}
Single photon sources have important applications in quantum information protocols such as quantum key distributions~\cite{lucamarini2018overcoming}, quantum cryptography~\cite{gisin2002quantum}, quantum entanglement~\cite{volz2006observation} and optical quantum computing~\cite{o2007optical} etc. Photon blockade (PB) is a key technique and a major contender ~\cite{lounis2005single} among various single photon source schemes. In a conventional PB, the energy-level anharmonicity (ELA) implies that the absorption of a photon necessarily blocks the corresponding transition, preventing transmission of later arriving photons at the same frequency. This scheme can result in non-classical statistical characteristics of photons, yielding sub-Poissonian statistics and photon anti-bunching behavior~\cite{agarwal2012quantum} that are essential for applications in the field of quantum information science~\cite{kyriienko2014tunable}.

Cavity quantum electrodynamics (QED) system is a typical physical platform to realize the PB phenomenon, where a two-level qubit is strongly coupled to the cavity mode. The corresponding anharmonic energy intervals by cavity interaction reject all but a single photon of a specific driving field frequency, which may lead to anti-bunched photons. This ELA-induced PB requires strong coupling qubit-cavity strengths~\cite{Schneebeli2008} and is widely known as the conventional photon blockade (CPB)~\cite{imamoglu1997strongly}. To date, ELA-based PB has been experimentally and theoretically studied in various systems~\cite{Kuhn2002,birnbaum2005photon,faraon2008coherent,ridolfo2012photon,Muller2015,Sazquez2018}, including atom-cavity QED~\cite{PhysRevLett.118.133604, PhysRevA.95.063842, PhysRevA.99.053850}, optomechanical systems~\cite{PhysRevLett.107.063601,PhysRevA.88.023853,PhysRevA.96.013861}, circuit QED systems~\cite{PhysRevLett.107.053602, PhysRevLett.106.243601,PhysRevA.89.043818}, Kerr-nonlinearity systems~\cite{PhysRevA.87.023809,PhysRevA.90.013839,PhysRevLett.121.153601}, and so on.

Quantum destructive interference effect (QDI) \cite{Deng2006} is a widely recognized optical excitation protocol where interference between multiple optical excitation pathways can lead to strong enhancement/suppression of selected optical transitions and effects such as ac Start shift suppression, ionization suppression, wave-mixing channel enhancement/suppression, and resonant enhanced ionization spectroscopy etc. It has also been used to realize PB and the corresponding scheme is referred to as the unconventional photon blockade (UPB)~\cite{Liew2011,Flayac2017}. Essentially, the QDI-induced PB is originated from a QDI effect involving two (or more) different excitation pathways~\cite{Bamba2011}, resulting in elimination of two-photon excitation. Since strong coupling strengths are not required for achieving QDI-induced PB~\cite{snijders2016purification}, this scheme has received great attention and it has been proposed for many quantum systems including multi-level atom-cavity QED system~\cite{tang2019strong}, quantum dots~\cite{ tang2015quantum}, third-order nonlinearity scheme~\cite{ferretti2013optimal,Shen2015}, optical parametric amplifier scheme~\cite{sarma2017quantum}, optomechanical device~\cite{wang2015tunable,li2019nonreciprocal,xu2016phonon}, non-Markovian system~\cite{PhysRevA.98.023856}, two-emitter-cavity~\cite{radulaski2017photon} and Jaynes-Cummings model~\cite{Bamba2011, Majumdar2012, liang2018photon}. Recently, the QDI-induced PB has been demonstrated in superconducting QED systems~\cite{Vaneph2018, Snijders2018}. 

Although the ELA-induced PB have been demonstrated in several systems it generally cannot yield very small second-order correlation function $g^{(2)}(0)$ which is a standard bench mark criterion for the performance of a PB. This is because in ELA-induced PB generally requires strong coupling where two- and perhaps multi-photon excitation cannot be simultaneously avoided. On the other hand, a QDI-induced PB has the advantage of being able to achieve extremely strong antibunching behavior which necessarily results in ideal $g^{(2)}(0)\rightarrow0$. However, because QDI generally requires the driving field intensity be weak enough to prevent multiphoton excitation pathways from becoming dominant the number of generated photon in the cavity is therefore very small, making experimental detection difficult. Clearly, the strong coupling requirement for ELA scheme and the weak coupling condition for QDI scheme are mutually exclusive and present an obstacle for simultaneously achieving large photon number and exceedingly small $g^{(2)}(0)$.

In this work, we propose a two-qubit (e.g., two two-level atoms) single-cavity system that combines the advantages of ELA scheme and QDI scheme for novel PB performance.  We study a hybrid PB scheme in which an ELA-based mechanism and a QDI-based mechanism act cooperatively in a two-qubit cavity QED system with dipole-dipole interaction (DDI). In this configuration a two-qubit system is coherently ``locked" by a cavity mode, effectively making it a ``super-qubit" system with two ``internal" excitation pathways. Here, each pathway involves one qubit in a coherently locked and driven two-qubit-one-cavity system. This new scheme mimics the diamond excitation scheme (therefore also referred to as a diamond PB or DPB) in nonlinear optics where robust QDI effect can play a dominant role~\cite{Deng2006}. We find that the excitation pathways via two individual but locked qubits are indistinct when the qubit resonance frequency is the same as that of the cavity mode, yielding the constructive interference. However, when the qubit resonance frequency is shifted from the cavity mode frequency, the two excitation pathways become distinct, yielding a destructive interference. In contrary to UPBs where the destructive interference is formed by different excitation pathways in the same single qubit, the proposed DPB scheme is independent of coupling strength. This implies that in the presence of the DDI both the ELA and QDI induced PBs, which are the two core elements of the DPB, can be achieved at the same coupling strength for the probe of same frequency. This opens the possibility for a hybrid and extremely strong PB effect that yields a well-detectable mean photon number, the virtual of an ELA-PB, and yet extremely small $g^{(2)}(0)$ which is the virtual of a QDI-PB.

\section{Model System}
\begin{figure}[htbp]
	\centering
	\includegraphics[width=\linewidth]{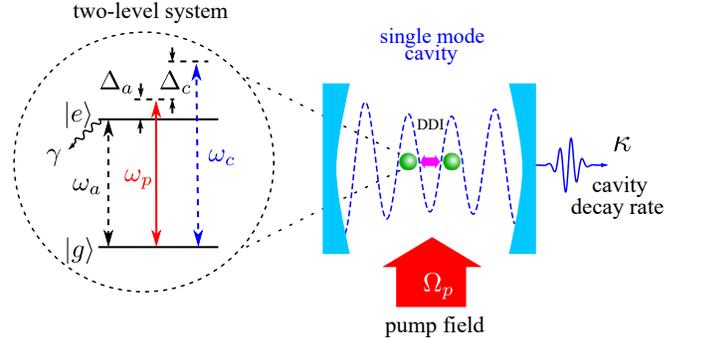}
	\caption{(Color online) The sketch of two qubits cavity QED system with different cavity mode frequency $\omega_c$ and qubit resonant frequency $\omega_a$. A pump field $\Omega_p$ couples the qubit ground state $|g\rangle$ and excited state $|e\rangle$ with the angular frequency $\omega_p$. $\gamma$ and $\kappa$ denote the qubit decay rate and the cavity decay rate, respectively. Here, DDI represents the dipole-dipole interaction when two qubits are close enough.}~\label{fig:model}
\end{figure}
We consider two identical two-level qubit with resonant frequency $\omega_a$ and are embedded inside a single mode cavity with resonant frequency $\omega_{c}$. The ground (excited) state is labeled as $|g\rangle$ ($|e\rangle$), respectively, and the positions of two qubits are given by $\mathbf{x}_j\ (j=1-2)$ (Fig.~\ref{fig:model}). The two qubits are coherently driven by a pump field with angular frequency $\omega_p$ and Rabi frequency $\Omega_p$. The dipole-dipole interaction (DDI) plays an important role to the statistical properties of the cavity field when $|\mathbf{x}_1-\mathbf{x}_2|<\lambda_a\equiv2\pi c/\omega_a$ and it can be safely neglected if the two qubits are well separated. Assuming that these two qubits have the same qubit-cavity coupling strength $g$ and using rotating wave approximation, the Hamiltonian in the presence of DDI can be expressed as
\begin{eqnarray}\label{eq:H}
H&=&\sum_{j=1}^2-\Delta_a\sigma_{j}^{\dag}\sigma_{j}-\Delta_c a^\dagger a+\sum_{j=1}^2g(a^\dagger\sigma_{j}+a\sigma_{j}^{\dag})\nonumber\\
&&+J(\sigma_{1}\sigma_{2}^{\dag}+\sigma_{2}\sigma_{1}^{\dag})
+\sum_{j=1}^2\Omega_{\rm p}(\sigma_{j}+\sigma_{j}^{\dag}),
\end{eqnarray}
where $\Delta_c=\omega_p-\omega_c$ and $\Delta_ a=\omega_p-\omega_a$ are the cavity and qubit frequency detunings, respectively. Here, $a\ (a^\dagger)$ is the cavity field annihilation (creation) operator, and $\sigma_{j}=|g\rangle_{j}\langle e|$ is the lowering operators of the j-th qubit. The DDI strength $J$ between two qubits is given by~\cite{ficek2002entangled}
\begin{eqnarray}\label{eq:J}
J&=&\dfrac{3\gamma}{4}\left\{-(1-\cos^2\theta)\frac{\cos(\mathbf{k}\cdot\mathbf{d})}{\mathbf{k}\cdot\mathbf{d}}\right.\nonumber\\
&&\left.+(1-3\cos^2\theta)\left[\dfrac{\sin(\mathbf{k}\cdot\mathbf{d})}{(\mathbf{k}\cdot\mathbf{d})^2}+\dfrac{\cos(\mathbf{k}\cdot\mathbf{d})}{(\mathbf{k}\cdot\mathbf{d})^3}\right]\right\},
\end{eqnarray}
where $\theta$ is the angle between the qubit dipole moment $\boldsymbol{\mu}$ and the distance vector $\mathbf{d}=\mathbf{x}_2-\mathbf{x}_1$. $\gamma$ is the qubit decay rate in free space and $k=\omega_a/c$ is the wave vector for the qubit resonant frequency $\omega_a$. 

The dynamics of this two-quibit coherently driven system is described by the master equation~\cite{agarwal2012quantum}:
\begin{eqnarray}\label{eq:master}
\frac{\partial{\rho}}{\partial{t}}=-i[H,\rho]+\mathcal{L}_\kappa[\rho]+\mathcal{L}_\gamma[\rho]+\mathcal{L}_{\gamma'}[\rho]
\end{eqnarray}
where $\rho$ is the system density matrix, $\mathcal{L}_\kappa[\rho]=\kappa(2a\rho a^\dagger-a^\dagger a\rho-\rho a^\dagger a)$ describes the cavity leakage with rate $\kappa$, and $\mathcal{L}_\gamma[\rho]=\sum_{j=1}^{2}\gamma(2\sigma_{j}\rho\sigma_{j}^\dagger-\sigma_{j}^\dagger\sigma_{j}\rho-\rho\sigma_{j}^\dagger\sigma_{j})$ indicates free space damping of the j-th qubit with rate $\gamma$. The last term $\mathcal{L}_{\gamma'}[\rho]=\gamma'\sum_{i\neq j}(2\sigma_{i}\rho\sigma_{j}^\dagger-\sigma_{i}^\dagger\sigma_{j}\rho-\rho\sigma_{i}^\dagger\sigma_{j})$
describes the collective damping resulting from the mutual exchange of spontaneously emitted photons through the common reservoir, and the collective emission rate is given by~\cite{PhysRevA.84.013831}
\begin{eqnarray}\label{eq:col}
\gamma'&=&\dfrac{3\gamma}{2}\left[(1-\cos^2\theta)\frac{\sin(\mathbf{k}\cdot\mathbf{d})}{\mathbf{k}\cdot\mathbf{d}}\right.\nonumber\\
&&\left.+(1-3\cos^2\theta)(\dfrac{\cos(\mathbf{k}\cdot\mathbf{d})}{(\mathbf{k}\cdot\mathbf{d})^2}-\dfrac{\sin(\mathbf{k}\cdot\mathbf{d})}{(\mathbf{k}\cdot\mathbf{d})^3})\right].
\end{eqnarray}

\section{Energy level anharmonicity and quantum interference induced PBs in the absence of DDI}
\begin{figure*}[htbp]
	\centering
	\includegraphics[width=\linewidth]{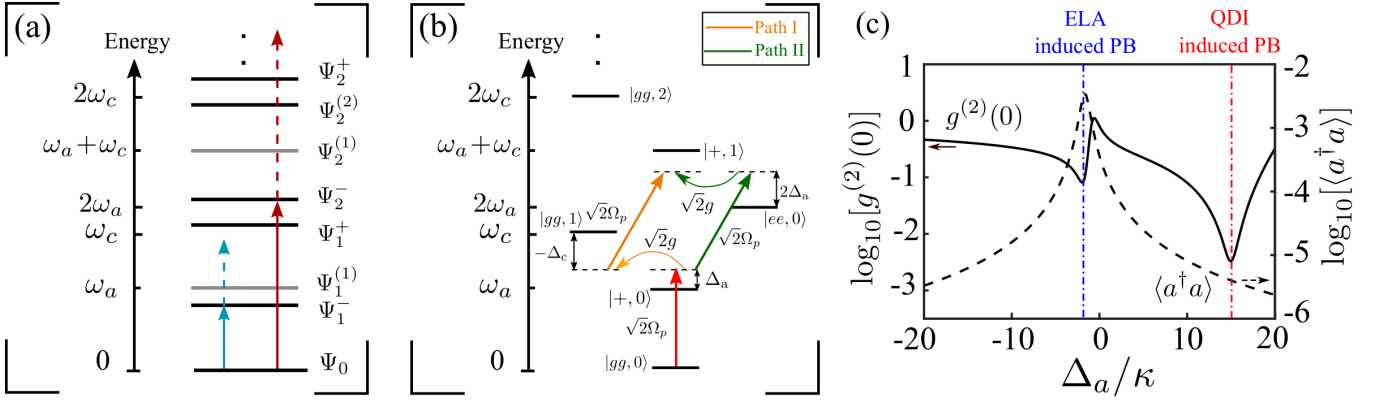}
	\caption{(Color online)  Panels (a) and (b) show the ladder-type energy structure and the diamond transition pathways for the ELA-based and QDI-induced PBs, respectively. Panel (c): the equal-time second-order correlation function $g^{(2)}(0)$ (solid curve) and mean photon number $\langle a^\dagger a\rangle$ (dashed curve) as a function of the normalized detuning $\Delta_a/\kappa$. Here, we chose $J=0$, $\Delta_c=-30\kappa$, $g=5\kappa$, $\gamma=\kappa$ and $\Omega_p=0.1\kappa$.}~\label{fig:no-ddi}
\end{figure*}
We first consider the case where the DDI is absent, i.e., two qubits are well separated with $kd\gg1$. In this case, using collective states, i.e., $|gg,n\rangle$, $|\pm,n\rangle\equiv(|eg,n\rangle+|ge,n\rangle)/\sqrt{2}$ and $|ee,n\rangle$, one can rewrite the Hamiltonian (1) to give $H=\Delta_aS^\dag S+\Delta_ca^\dag a+2g(aS^\dag+a^\dag S)+\Omega_p(S+S^\dag)$ with collective operator $S=(\sigma_1+\sigma_2)/\sqrt{2}$. In the strong coupling regime, it is convenient to describe the system by using the dressed state representation [see Fig.~\ref{fig:no-ddi}(a) and supplementary material]. In $n$-photon space the eigenstates form ladder-type doublets with unevenly separated energy levels. When the pump field was tuned to one of the states of the lowest doublet, e.g., $2g^2=\Delta_a\Delta_c$, the absorption of a second photon of the pump field will be blocked due to the large energy mismatch resulted from energy level anharmonicity. Figure~\ref{fig:no-ddi}(c) depicts the second order correlation function $g^{(2)}(0)=\langle a^\dag a^\dag a a \rangle/(\langle a^\dag a \rangle)^2<1$ for this DDI-neglected DPB scheme. Here, we choose system parameters as $\Delta_c=-30\kappa$, $g=5\kappa$, $\gamma=\kappa$, and $\Omega_p=0.1\kappa$. 

Figure~\ref{fig:no-ddi}(c) exhibits two PBs originated from different mechanisms. The blue dash-dotted line indicates the detuning $\Delta_a \approx -1.6\kappa$ for realizing an ELA-induced PB operation whereas the red dash-dotted line shows the location of the QDI-induced PB operation. The latter exhibits near two orders of magnitude smaller $g^{(2)}(0)$, a virtue of the QDI-based PB operation.  However, the mean photon number only peaks at the location of ELA-based PB operation. The three-order of magnitude difference in mean photon number demonstrates the significant experimental detection advantage of ELA scheme over the QDI scheme. 

To elucidate two separately located PBs of different origins we examine the two-qubit-cavity coupling without using the dressed state picture. Figure ~\ref{fig:no-ddi}(b) exhibits two pathways for two-photon excitation of the state $|+,1\rangle$ after direct one-photon excitation of the state $|+,0\rangle$. One path is indicated by straight and curved yellow arrows representing $|+,0\rangle\overset{\sqrt{2}g}{\rightarrow}|gg,1\rangle\overset{\sqrt{2}\Omega_p}{\rightarrow}|+,1\rangle$ and the other path is indicated by straight and curved green arrows representing $|+,0\rangle\overset{\sqrt{2}\Omega_p}{\rightarrow}|ee,0\rangle\overset{\sqrt{2}g}{\rightarrow}|+,1\rangle$, respectively. When $\Delta_a=\Delta_c$, these two excitation pathways are indistinguishable so that the two photon excitation can be enhanced due to the constructive interference. However, when $\Delta_a\neq\Delta_c$, these two excitation pathways are distinct and form a destructive interference which leads to the blockade of two-photon excitation of state $|+,1\rangle$. This is the essence of QDI-induced PB effect. Using the amplitude equations, one can show that the condition for achieving this QDI-induced PB is $\Delta_c=-2\Delta_a$ which is {\it independent} of the coupling strength, a virtue which can be used to achieve a novel PB operation with large mean photon number and yet extremely small $g^{(2)}(0)$ [see discussion below and supplementary material]. 

We note that the such a quantum interference by excitation pathways does not exist in a single qubit cavity-driven QED system. In the single qubit case, the two-photon state $|g,2\rangle$ can only be excited via a single pathway, i.e., $|g,0\rangle\overset{\Omega_p}{\rightarrow}|e,0\rangle\overset{g}{\rightarrow}|g,1\rangle\overset{\Omega_p}{\rightarrow}|e,1\rangle\overset{g}{\rightarrow}|g,2\rangle$. We also note that the QDI-induce PB in the two qubits cavity-driven QED system described in this work is different from those reported in literature~\cite{Flayac2017,Vaneph2018, Snijders2018}. In our DPB system, the condition for realizing the QDI-induced PB is insensitive to the coupling strength and this indicates that the scheme is more robust and immune to field-related fluctuations in applications. 

\section{Dipole-dipole interaction (DDI) induced Strong PB effect}
\begin{figure}[htbp]
	\centering
	\includegraphics[width=\linewidth]{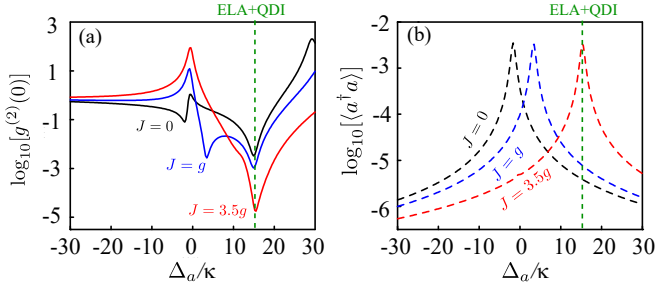}
	\caption{(Color online) Plots of the second-order correlation function $g^{(2)}(0)$ [panel (a)] and the mean photon number $\langle a^\dagger a\rangle$ [panel (b)] as a function of the normalized detuning $\Delta_a/\kappa$ with the DDI strength $J=0$ (black curves), $g$ (blue curves) and $3.5g$ (red curves), respectively. Other system parameters are the same as those used in Fig.~2(c).}\label{fig:g2-2d}
\end{figure}
While Fig.~\ref{fig:no-ddi}(c) exhibits virtues of ELA-scheme and QDI-scheme these features occur at different PB operational frequencies and therefore this type of DPB is not useful. We naturally ask if the ELA-scheme and QDI-scheme can be combined with an additional interaction control mechanism so that the advantages of both schemes can be maximized at the same PB frequency. As we show below that the combination of DPB and DDI can indeed achieve this goal, resulting in a hybrid DPB that preserves the virtue of large mean photon number and extremely small $g^{(2)}(0)$. 

In the presence of DDI, the state $|+,n\rangle$ is shifted by an amount of $J$ which characterizes the DDI strength. Consequently, the condition for achieving ELA-induced PB becomes $2g^2=\Delta_c(\Delta_a-J)$. However, for the QDI-induced PB because both $|+,0\rangle$ and $|+,1\rangle$ states are shifted by the same amount of $J$ in the same direction, the condition for realizing the QDI-induced PB remains the same. It is this differential change that provides a tunability that can lead to overlap of the operational frequencies of both schemes. This desired operation regime can be achieved by making $\Delta_c=-2\Delta_a$ and which results in a very strong PB phenomenon as shown in Fig.~\ref{fig:g2-2d}(a) and \ref{fig:g2-2d}(b). Here, an order of magnitude mean photon number increase and more than four orders of magnitude reduction of $g^{(2)}(0)$ are achieved simultaneously at the same PB operational frequency of $\Delta_a/\kappa=15$.  [see the green dashed lines and the black curves are the same as those shown in Fig.~\ref{fig:no-ddi}(c)]. %The system parameters are chosen as $\Omega_p=0.1\kappa$, $g=10\kappa$, $\gamma=\kappa$, and $\Delta_c=-20\kappa$. 

\begin{figure}[htbp]
	\centering
	\includegraphics[width=\linewidth]{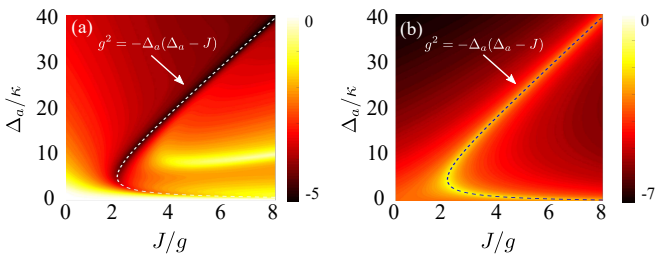}
	\caption{(Color online) Plots of the second-order correlation function $g^{(2)}(0)$ [panel (a)] and the mean photon number $\langle a^\dag a\rangle$ [panel (b)] as functions of the DDI strength $J/g$ and  detuning $\Delta_a/\kappa$ by setting $\Delta_c=-2\Delta_a$ and $g=5\kappa$. Other system parameters are the same as those used in Fig.~\ref{fig:g2-2d}. The  dashed curves denote the optimal condition $g^2=-\Delta_a(\Delta_a-J)$.}\label{fig:J-3d}
\end{figure}
In Fig.~\ref{fig:J-3d}, we plot the second-order correlation function $g^{(2)}(0)$ [panel (a)] and the mean photon number $\langle a^\dag a\rangle$ [panel (b)] as functions of the DDI strength $J/g$ and detuning $\Delta_a/\kappa$ by setting $\Delta_c=-2\Delta_a$ (other system parameters are the same as those used in Fig.~\ref{fig:g2-2d}). The condition for frequency matched operation $g^2=-\Delta_a(\Delta_a-J)$ is indicated by the dashed curves. It can be shown mathematically that $J\geq 2g$ is necessary for a PB operation with minimum $g^{(2)}(0)$. It is also interesting to note that the DDI plays the role similar to the tunneling between different cites in the Bose-Hubbard Hamiltonian and the detuning dependency of the energy spectrum has similarities like interband transitions in solid state materials, where the crossing to the maximum PB effect ``mimics" the crossing of different band structures (see supplementary material).

\begin{figure}[htbp]
	\centering
	\includegraphics[width=\linewidth]{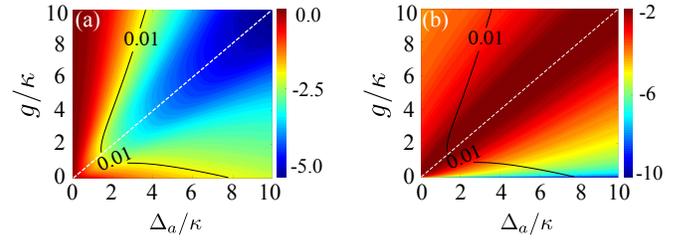}
	\caption{(Color online) The second-order correlation function $g^{(2)}(0)$ [panel (a)] and the mean photon number $\langle a^\dag a\rangle$ [panel (b)] versus the detuning $\Delta_a/\kappa$ and coupling strength $g/\kappa$ with $\Delta_c=-2\Delta_a$ and $J=2g$. The white dashed line corresponds to the condition $g=\Delta_a$, while the black solid curves denote $g^{(2)}(0)=0.01$.}\label{fig:g-3d}
\end{figure}
The influence of qubit-cavity coupling strength to the PB is shown in Fig.\ref{fig:g-3d}(a) and (b) where we show the contour plots of $\log_{10}[g^{(2)}(0)]$ and $\log_{10}[\langle a^\dagger a\rangle]$ as functions of the normalized detuning $\Delta_a/\kappa$ and atom-cavity coupling strength $g/\kappa$ by taking $\Delta_c=-2\Delta_a$ and $J=2g$. The white dashed lines denote the optimal condition of PB operation, i.e., $g=\Delta_a$. With this optimal condition, the $g^{(2)}(0)$ of the PB can be improved by increasing the atom-cavity coupling strength [see panel (a)].  In Fig.~\ref{fig:g-3d}(b) the mean photon number is always at its maximum but the parameter space for reaching this optimal number increases. It is worthy to point out that the DPB scheme has a broader range of parameters to realize $g^{(2)}(0)<0.01$ than the single atom UPB scheme. [see Fig.~4(a) and Fig.~5(a)].

\begin{figure}[htbp]
	\centering
	\includegraphics[width=\linewidth]{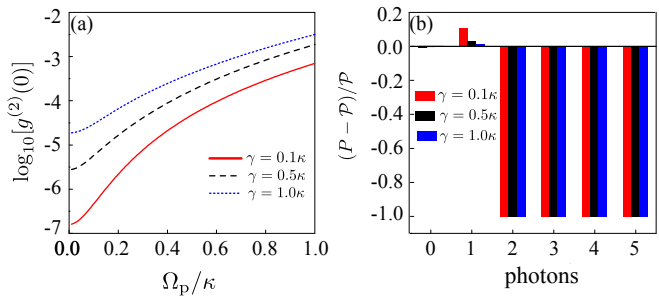}
	\caption{
		(Color online) Panel (a): the second-order correlation function versus the normalize pump field Rabi frequency $\Omega_p/\kappa$ with spontaneous decay rate $\gamma/2\pi=0.1\kappa$  (red solid curves), $0.5\kappa$ (black dashed curves) and $\kappa$ (blue dotted curves), respectively. Panel (b): the plot of $(P-{\cal P})/{\cal P}$ as a function of the photon number in Fock states with $\Omega_p=0.2\kappa$. Here, the system parameters are given by $g=2\kappa$, and $J=\Delta_c=-2\Delta_a=2g$.
	}\label{fig:elta-3d}
\end{figure}
Driving field Rabi frequency and atomic decay rate also affect the PB effect. As show in Fig.~\ref{fig:elta-3d}(a), although $g^{(2)}(0)$ increases as the driving filed Rabi frequency $\Omega_p$ increases strong PB phenomenon (defined as $g^{(2)}<0.01$) still can be achieved. Moreover, the smaller the atomic spontaneous decay rate the stronger the PB effect becomes. In Fig.~\ref{fig:elta-3d}(b), we show the quantum statistic properties of the cavity field by calculating the relative deviations of the cavity photon distribution P($n$) from the Poisson distribution $\mathcal{P}(n)$ with the same photon number, i.e., $[{\rm P}(n)-{\cal P}(n)]/{\cal P}(n)$. It is clear that the probabilities for detecting $n\geq2$ photons are strongly suppressed. %Here, we choose $\Omega_p=0.2\kappa$ and other system parameters are the same as those used in Fig.~6a.

\section{conclusion}
In summary, we have investigated photon blockade effect using a two-qubit cavity driven QED system with combined ELA/QDI-scheme and atom-atom dipole-dipole interaction included. In the absence of DDI, we show that the QDI-induced photon blockade can be achieved only if the qubit resonance frequency is different from the cavity mode frequency and the mean photon number is low. Using the amplitude method, we derived conditions for the ELA-based CPB and QDI-based UPB. We showed that the condition for QDI-induced PB depends only on the pump field frequency and the effect is insensitive to the qubit-cavity coupling strength. We also showed that extremely strong PB effect with much improved mean photon numbers can be realized using two qubits dipole-dipole interaction. Our work provides theoretical foundation for possible experimental demonstration of this diamond-scheme for highly efficient PB operation. The implementation of the protocol is potentially demonstrable using various quantum systems such as semiconductor quantum dot or quantum well cavity QED system~\cite{bayer2001coupling,cox2012dipole}, Rydberg atom-cavity QED system~\cite{saffman2005analysis}, and circuit cavity QED system~\cite{quijandria2018pt}. It may lead to a new type of hybrid single photon sources for quantum information processing and communication. 

\begin{acknowledgments}
This work was supported by the National Key Basic Research Special Foundation (Grant No.2016YFA0302800); the Shanghai Science and Technology Committee (Grants No.18JC1410900); the National Nature Science Foundation (Grant No.11774262). 
%; the Natural Science Foundation of Anhui Province (Grant No. 1608085QA23). Natural Science Foundation of Anhui Provincial Education Department (Grant No.KJ2018JD20).
\end{acknowledgments}

\bibliography{ref-ddi}

\begin{thebibliography}{52}%
\makeatletter
\providecommand \@ifxundefined [1]{%
 \@ifx{#1\undefined}
}%
\providecommand \@ifnum [1]{%
 \ifnum #1\expandafter \@firstoftwo
 \else \expandafter \@secondoftwo
 \fi
}%
\providecommand \@ifx [1]{%
 \ifx #1\expandafter \@firstoftwo
 \else \expandafter \@secondoftwo
 \fi
}%
\providecommand \natexlab [1]{#1}%
\providecommand \enquote  [1]{``#1''}%
\providecommand \bibnamefont  [1]{#1}%
\providecommand \bibfnamefont [1]{#1}%
\providecommand \citenamefont [1]{#1}%
\providecommand \href@noop [0]{\@secondoftwo}%
\providecommand \href [0]{\begingroup \@sanitize@url \@href}%
\providecommand \@href[1]{\@@startlink{#1}\@@href}%
\providecommand \@@href[1]{\endgroup#1\@@endlink}%
\providecommand \@sanitize@url [0]{\catcode `\\12\catcode `\$12\catcode
  `\&12\catcode `\#12\catcode `\^12\catcode `\_12\catcode `\%12\relax}%
\providecommand \@@startlink[1]{}%
\providecommand \@@endlink[0]{}%
\providecommand \url  [0]{\begingroup\@sanitize@url \@url }%
\providecommand \@url [1]{\endgroup\@href {#1}{\urlprefix }}%
\providecommand \urlprefix  [0]{URL }%
\providecommand \Eprint [0]{\href }%
\providecommand \doibase [0]{http://dx.doi.org/}%
\providecommand \selectlanguage [0]{\@gobble}%
\providecommand \bibinfo  [0]{\@secondoftwo}%
\providecommand \bibfield  [0]{\@secondoftwo}%
\providecommand \translation [1]{[#1]}%
\providecommand \BibitemOpen [0]{}%
\providecommand \bibitemStop [0]{}%
\providecommand \bibitemNoStop [0]{.\EOS\space}%
\providecommand \EOS [0]{\spacefactor3000\relax}%
\providecommand \BibitemShut  [1]{\csname bibitem#1\endcsname}%
\let\auto@bib@innerbib\@empty
%</preamble>
\bibitem [{\citenamefont {Lucamarini}\ \emph {et~al.}(2018)\citenamefont
  {Lucamarini}, \citenamefont {Yuan}, \citenamefont {Dynes},\ and\
  \citenamefont {Shields}}]{lucamarini2018overcoming}%
  \BibitemOpen
  \bibfield  {author} {\bibinfo {author} {\bibfnamefont {Marco}\ \bibnamefont
  {Lucamarini}}, \bibinfo {author} {\bibfnamefont {Zhiliang~L}\ \bibnamefont
  {Yuan}}, \bibinfo {author} {\bibfnamefont {James~F}\ \bibnamefont {Dynes}}, \
  and\ \bibinfo {author} {\bibfnamefont {Andrew~J}\ \bibnamefont {Shields}},\
  }\bibfield  {title} {\enquote {\bibinfo {title} {Overcoming the
  rate--distance limit of quantum key distribution without quantum
  repeaters},}\ }\href@noop {} {\bibfield  {journal} {\bibinfo  {journal}
  {Nature}\ }\textbf {\bibinfo {volume} {557}},\ \bibinfo {pages} {400}
  (\bibinfo {year} {2018})}\BibitemShut {NoStop}%
\bibitem [{\citenamefont {Gisin}\ \emph {et~al.}(2002)\citenamefont {Gisin},
  \citenamefont {Ribordy}, \citenamefont {Tittel},\ and\ \citenamefont
  {Zbinden}}]{gisin2002quantum}%
  \BibitemOpen
  \bibfield  {author} {\bibinfo {author} {\bibfnamefont {Nicolas}\ \bibnamefont
  {Gisin}}, \bibinfo {author} {\bibfnamefont {Gr{\'e}goire}\ \bibnamefont
  {Ribordy}}, \bibinfo {author} {\bibfnamefont {Wolfgang}\ \bibnamefont
  {Tittel}}, \ and\ \bibinfo {author} {\bibfnamefont {Hugo}\ \bibnamefont
  {Zbinden}},\ }\bibfield  {title} {\enquote {\bibinfo {title} {Quantum
  cryptography},}\ }\href@noop {} {\bibfield  {journal} {\bibinfo  {journal}
  {Rev. Mod. Phys.}\ }\textbf {\bibinfo {volume} {74}},\ \bibinfo {pages} {145}
  (\bibinfo {year} {2002})}\BibitemShut {NoStop}%
\bibitem [{\citenamefont {Volz}\ \emph {et~al.}(2006)\citenamefont {Volz},
  \citenamefont {Weber}, \citenamefont {Schlenk}, \citenamefont {Rosenfeld},
  \citenamefont {Vrana}, \citenamefont {Saucke}, \citenamefont {Kurtsiefer},\
  and\ \citenamefont {Weinfurter}}]{volz2006observation}%
  \BibitemOpen
  \bibfield  {author} {\bibinfo {author} {\bibfnamefont {J{\"u}rgen}\
  \bibnamefont {Volz}}, \bibinfo {author} {\bibfnamefont {Markus}\ \bibnamefont
  {Weber}}, \bibinfo {author} {\bibfnamefont {Daniel}\ \bibnamefont {Schlenk}},
  \bibinfo {author} {\bibfnamefont {Wenjamin}\ \bibnamefont {Rosenfeld}},
  \bibinfo {author} {\bibfnamefont {Johannes}\ \bibnamefont {Vrana}}, \bibinfo
  {author} {\bibfnamefont {Karen}\ \bibnamefont {Saucke}}, \bibinfo {author}
  {\bibfnamefont {Christian}\ \bibnamefont {Kurtsiefer}}, \ and\ \bibinfo
  {author} {\bibfnamefont {Harald}\ \bibnamefont {Weinfurter}},\ }\bibfield
  {title} {\enquote {\bibinfo {title} {Observation of entanglement of a single
  photon with a trapped atom},}\ }\href@noop {} {\bibfield  {journal} {\bibinfo
   {journal} {Phys. Rev. Lett.}\ }\textbf {\bibinfo {volume} {96}},\ \bibinfo
  {pages} {030404} (\bibinfo {year} {2006})}\BibitemShut {NoStop}%
\bibitem [{\citenamefont {O'brien}(2007)}]{o2007optical}%
  \BibitemOpen
  \bibfield  {author} {\bibinfo {author} {\bibfnamefont {Jeremy~L}\
  \bibnamefont {O'brien}},\ }\bibfield  {title} {\enquote {\bibinfo {title}
  {Optical quantum computing},}\ }\href@noop {} {\bibfield  {journal} {\bibinfo
   {journal} {Science}\ }\textbf {\bibinfo {volume} {318}},\ \bibinfo {pages}
  {1567--1570} (\bibinfo {year} {2007})}\BibitemShut {NoStop}%
\bibitem [{\citenamefont {Lounis}\ and\ \citenamefont
  {Orrit}(2005)}]{lounis2005single}%
  \BibitemOpen
  \bibfield  {author} {\bibinfo {author} {\bibfnamefont {Brahim}\ \bibnamefont
  {Lounis}}\ and\ \bibinfo {author} {\bibfnamefont {Michel}\ \bibnamefont
  {Orrit}},\ }\bibfield  {title} {\enquote {\bibinfo {title} {Single-photon
  sources},}\ }\href@noop {} {\bibfield  {journal} {\bibinfo  {journal}
  {Reports on Progress in Physics}\ }\textbf {\bibinfo {volume} {68}},\
  \bibinfo {pages} {1129} (\bibinfo {year} {2005})}\BibitemShut {NoStop}%
\bibitem [{\citenamefont {Agarwal}(2013)}]{agarwal2012quantum}%
  \BibitemOpen
  \bibfield  {author} {\bibinfo {author} {\bibfnamefont {G.~S.}\ \bibnamefont
  {Agarwal}},\ }\href@noop {} {\emph {\bibinfo {title} {Quantum optics}}}\
  (\bibinfo  {publisher} {CambridgeUniversity Press, New York},\ \bibinfo
  {year} {2013})\BibitemShut {NoStop}%
\bibitem [{\citenamefont {Kyriienko}\ \emph {et~al.}(2014)\citenamefont
  {Kyriienko}, \citenamefont {Shelykh},\ and\ \citenamefont
  {Liew}}]{kyriienko2014tunable}%
  \BibitemOpen
  \bibfield  {author} {\bibinfo {author} {\bibfnamefont {Oleksandr}\
  \bibnamefont {Kyriienko}}, \bibinfo {author} {\bibfnamefont {Ivan~A}\
  \bibnamefont {Shelykh}}, \ and\ \bibinfo {author} {\bibfnamefont {Timothy
  Chi~Hin}\ \bibnamefont {Liew}},\ }\bibfield  {title} {\enquote {\bibinfo
  {title} {Tunable single-photon emission from dipolaritons},}\ }\href@noop {}
  {\bibfield  {journal} {\bibinfo  {journal} {Phys. Rev. A}\ }\textbf {\bibinfo
  {volume} {90}},\ \bibinfo {pages} {033807} (\bibinfo {year}
  {2014})}\BibitemShut {NoStop}%
\bibitem [{\citenamefont {Schneebeli}\ \emph {et~al.}(2008)\citenamefont
  {Schneebeli}, \citenamefont {Kira},\ and\ \citenamefont
  {Koch}}]{Schneebeli2008}%
  \BibitemOpen
  \bibfield  {author} {\bibinfo {author} {\bibfnamefont {L.}~\bibnamefont
  {Schneebeli}}, \bibinfo {author} {\bibfnamefont {M.}~\bibnamefont {Kira}}, \
  and\ \bibinfo {author} {\bibfnamefont {S.~W.}\ \bibnamefont {Koch}},\
  }\bibfield  {title} {\enquote {\bibinfo {title} {Characterization of strong
  light-matter coupling in semiconductor quantum-dot microcavities via
  photon-statistics spectroscopy},}\ }\href {\doibase
  10.1103/PhysRevLett.101.097401} {\bibfield  {journal} {\bibinfo  {journal}
  {Phys. Rev. Lett.}\ }\textbf {\bibinfo {volume} {101}},\ \bibinfo {pages}
  {097401} (\bibinfo {year} {2008})}\BibitemShut {NoStop}%
\bibitem [{\citenamefont {Imamo\={g}lu}\ \emph {et~al.}(1997)\citenamefont
  {Imamo\={g}lu}, \citenamefont {Schmidt}, \citenamefont {Woods},\ and\
  \citenamefont {Deutsch}}]{imamoglu1997strongly}%
  \BibitemOpen
  \bibfield  {author} {\bibinfo {author} {\bibfnamefont {A}~\bibnamefont
  {Imamo\={g}lu}}, \bibinfo {author} {\bibfnamefont {Helmut}\ \bibnamefont
  {Schmidt}}, \bibinfo {author} {\bibfnamefont {Gareth}\ \bibnamefont {Woods}},
  \ and\ \bibinfo {author} {\bibfnamefont {Moshe}\ \bibnamefont {Deutsch}},\
  }\bibfield  {title} {\enquote {\bibinfo {title} {Strongly interacting photons
  in a nonlinear cavity},}\ }\href@noop {} {\bibfield  {journal} {\bibinfo
  {journal} {Phys. Rev. Lett.}\ }\textbf {\bibinfo {volume} {79}},\ \bibinfo
  {pages} {1467} (\bibinfo {year} {1997})}\BibitemShut {NoStop}%
\bibitem [{\citenamefont {Kuhn}\ \emph {et~al.}(2002)\citenamefont {Kuhn},
  \citenamefont {Hennrich},\ and\ \citenamefont {Rempe}}]{Kuhn2002}%
  \BibitemOpen
  \bibfield  {author} {\bibinfo {author} {\bibfnamefont {Axel}\ \bibnamefont
  {Kuhn}}, \bibinfo {author} {\bibfnamefont {Markus}\ \bibnamefont {Hennrich}},
  \ and\ \bibinfo {author} {\bibfnamefont {Gerhard}\ \bibnamefont {Rempe}},\
  }\bibfield  {title} {\enquote {\bibinfo {title} {Deterministic single-photon
  source for distributed quantum networking},}\ }\href {\doibase
  10.1103/PhysRevLett.89.067901} {\bibfield  {journal} {\bibinfo  {journal}
  {Phys. Rev. Lett.}\ }\textbf {\bibinfo {volume} {89}},\ \bibinfo {pages}
  {067901} (\bibinfo {year} {2002})}\BibitemShut {NoStop}%
\bibitem [{\citenamefont {Birnbaum}\ \emph {et~al.}(2005)\citenamefont
  {Birnbaum}, \citenamefont {Boca}, \citenamefont {Miller}, \citenamefont
  {Boozer}, \citenamefont {Northup},\ and\ \citenamefont
  {Kimble}}]{birnbaum2005photon}%
  \BibitemOpen
  \bibfield  {author} {\bibinfo {author} {\bibfnamefont {Kevin~M}\ \bibnamefont
  {Birnbaum}}, \bibinfo {author} {\bibfnamefont {Andreea}\ \bibnamefont
  {Boca}}, \bibinfo {author} {\bibfnamefont {Russell}\ \bibnamefont {Miller}},
  \bibinfo {author} {\bibfnamefont {Allen~D}\ \bibnamefont {Boozer}}, \bibinfo
  {author} {\bibfnamefont {Tracy~E}\ \bibnamefont {Northup}}, \ and\ \bibinfo
  {author} {\bibfnamefont {H~Jeff}\ \bibnamefont {Kimble}},\ }\bibfield
  {title} {\enquote {\bibinfo {title} {Photon blockade in an optical cavity
  with one trapped atom},}\ }\href@noop {} {\bibfield  {journal} {\bibinfo
  {journal} {Nature}\ }\textbf {\bibinfo {volume} {436}},\ \bibinfo {pages}
  {87} (\bibinfo {year} {2005})}\BibitemShut {NoStop}%
\bibitem [{\citenamefont {Faraon}\ \emph {et~al.}(2008)\citenamefont {Faraon},
  \citenamefont {Fushman}, \citenamefont {Englund}, \citenamefont {Stoltz},
  \citenamefont {Petroff},\ and\ \citenamefont
  {Vu{\v{c}}kovi{\'c}}}]{faraon2008coherent}%
  \BibitemOpen
  \bibfield  {author} {\bibinfo {author} {\bibfnamefont {Andrei}\ \bibnamefont
  {Faraon}}, \bibinfo {author} {\bibfnamefont {Ilya}\ \bibnamefont {Fushman}},
  \bibinfo {author} {\bibfnamefont {Dirk}\ \bibnamefont {Englund}}, \bibinfo
  {author} {\bibfnamefont {Nick}\ \bibnamefont {Stoltz}}, \bibinfo {author}
  {\bibfnamefont {Pierre}\ \bibnamefont {Petroff}}, \ and\ \bibinfo {author}
  {\bibfnamefont {Jelena}\ \bibnamefont {Vu{\v{c}}kovi{\'c}}},\ }\bibfield
  {title} {\enquote {\bibinfo {title} {Coherent generation of non-classical
  light on a chip via photon-induced tunnelling and blockade},}\ }\href@noop {}
  {\bibfield  {journal} {\bibinfo  {journal} {Nat. Phys.}\ }\textbf {\bibinfo
  {volume} {4}},\ \bibinfo {pages} {859} (\bibinfo {year} {2008})}\BibitemShut
  {NoStop}%
\bibitem [{\citenamefont {Ridolfo}\ \emph {et~al.}(2012)\citenamefont
  {Ridolfo}, \citenamefont {Leib}, \citenamefont {Savasta},\ and\ \citenamefont
  {Hartmann}}]{ridolfo2012photon}%
  \BibitemOpen
  \bibfield  {author} {\bibinfo {author} {\bibfnamefont {Alessandro}\
  \bibnamefont {Ridolfo}}, \bibinfo {author} {\bibfnamefont {Martin}\
  \bibnamefont {Leib}}, \bibinfo {author} {\bibfnamefont {Salvatore}\
  \bibnamefont {Savasta}}, \ and\ \bibinfo {author} {\bibfnamefont {Michael~J}\
  \bibnamefont {Hartmann}},\ }\bibfield  {title} {\enquote {\bibinfo {title}
  {Photon blockade in the ultrastrong coupling regime},}\ }\href@noop {}
  {\bibfield  {journal} {\bibinfo  {journal} {Phys. Rev. Lett.}\ }\textbf
  {\bibinfo {volume} {109}},\ \bibinfo {pages} {193602} (\bibinfo {year}
  {2012})}\BibitemShut {NoStop}%
\bibitem [{\citenamefont {M\"uller}\ \emph {et~al.}(2015)\citenamefont
  {M\"uller}, \citenamefont {Rundquist}, \citenamefont {Fischer}, \citenamefont
  {Sarmiento}, \citenamefont {Lagoudakis}, \citenamefont {Kelaita},
  \citenamefont {S\'anchez Mu\~noz}, \citenamefont {del Valle}, \citenamefont
  {Laussy},\ and\ \citenamefont {Vu\ifmmode \check{c}\else
  \v{c}\fi{}kovi\ifmmode~\acute{c}\else \'{c}\fi{}}}]{Muller2015}%
  \BibitemOpen
  \bibfield  {author} {\bibinfo {author} {\bibfnamefont {Kai}\ \bibnamefont
  {M\"uller}}, \bibinfo {author} {\bibfnamefont {Armand}\ \bibnamefont
  {Rundquist}}, \bibinfo {author} {\bibfnamefont {Kevin~A.}\ \bibnamefont
  {Fischer}}, \bibinfo {author} {\bibfnamefont {Tomas}\ \bibnamefont
  {Sarmiento}}, \bibinfo {author} {\bibfnamefont {Konstantinos~G.}\
  \bibnamefont {Lagoudakis}}, \bibinfo {author} {\bibfnamefont {Yousif~A.}\
  \bibnamefont {Kelaita}}, \bibinfo {author} {\bibfnamefont {Carlos}\
  \bibnamefont {S\'anchez Mu\~noz}}, \bibinfo {author} {\bibfnamefont {Elena}\
  \bibnamefont {del Valle}}, \bibinfo {author} {\bibfnamefont {Fabrice~P.}\
  \bibnamefont {Laussy}}, \ and\ \bibinfo {author} {\bibfnamefont {Jelena}\
  \bibnamefont {Vu\ifmmode \check{c}\else \v{c}\fi{}kovi\ifmmode~\acute{c}\else
  \'{c}\fi{}}},\ }\bibfield  {title} {\enquote {\bibinfo {title} {Coherent
  generation of nonclassical light on chip via detuned photon blockade},}\
  }\href {\doibase 10.1103/PhysRevLett.114.233601} {\bibfield  {journal}
  {\bibinfo  {journal} {Phys. Rev. Lett.}\ }\textbf {\bibinfo {volume} {114}},\
  \bibinfo {pages} {233601} (\bibinfo {year} {2015})}\BibitemShut {NoStop}%
\bibitem [{\citenamefont {S\'aez-Bl\'azquez}\ \emph {et~al.}(2018)\citenamefont
  {S\'aez-Bl\'azquez}, \citenamefont {Feist}, \citenamefont
  {Garc\'{\i}a-Vidal},\ and\ \citenamefont
  {Fern\'andez-Dom\'{\i}nguez}}]{Sazquez2018}%
  \BibitemOpen
  \bibfield  {author} {\bibinfo {author} {\bibfnamefont {R.}~\bibnamefont
  {S\'aez-Bl\'azquez}}, \bibinfo {author} {\bibfnamefont {J.}~\bibnamefont
  {Feist}}, \bibinfo {author} {\bibfnamefont {F.~J.}\ \bibnamefont
  {Garc\'{\i}a-Vidal}}, \ and\ \bibinfo {author} {\bibfnamefont {A.~I.}\
  \bibnamefont {Fern\'andez-Dom\'{\i}nguez}},\ }\bibfield  {title} {\enquote
  {\bibinfo {title} {Photon statistics in collective strong coupling:
  Nanocavities and microcavities},}\ }\href {\doibase
  10.1103/PhysRevA.98.013839} {\bibfield  {journal} {\bibinfo  {journal} {Phys.
  Rev. A}\ }\textbf {\bibinfo {volume} {98}},\ \bibinfo {pages} {013839}
  (\bibinfo {year} {2018})}\BibitemShut {NoStop}%
\bibitem [{\citenamefont {Hamsen}\ \emph {et~al.}(2017)\citenamefont {Hamsen},
  \citenamefont {Tolazzi}, \citenamefont {Wilk},\ and\ \citenamefont
  {Rempe}}]{PhysRevLett.118.133604}%
  \BibitemOpen
  \bibfield  {author} {\bibinfo {author} {\bibfnamefont {Christoph}\
  \bibnamefont {Hamsen}}, \bibinfo {author} {\bibfnamefont {Karl~Nicolas}\
  \bibnamefont {Tolazzi}}, \bibinfo {author} {\bibfnamefont {Tatjana}\
  \bibnamefont {Wilk}}, \ and\ \bibinfo {author} {\bibfnamefont {Gerhard}\
  \bibnamefont {Rempe}},\ }\bibfield  {title} {\enquote {\bibinfo {title}
  {Two-photon blockade in an atom-driven cavity qed system},}\ }\href {\doibase
  10.1103/PhysRevLett.118.133604} {\bibfield  {journal} {\bibinfo  {journal}
  {Phys. Rev. Lett.}\ }\textbf {\bibinfo {volume} {118}},\ \bibinfo {pages}
  {133604} (\bibinfo {year} {2017})}\BibitemShut {NoStop}%
\bibitem [{\citenamefont {Zhu}\ \emph {et~al.}(2017)\citenamefont {Zhu},
  \citenamefont {Yang},\ and\ \citenamefont {Agarwal}}]{PhysRevA.95.063842}%
  \BibitemOpen
  \bibfield  {author} {\bibinfo {author} {\bibfnamefont {C.~J.}\ \bibnamefont
  {Zhu}}, \bibinfo {author} {\bibfnamefont {Y.~P.}\ \bibnamefont {Yang}}, \
  and\ \bibinfo {author} {\bibfnamefont {G.~S.}\ \bibnamefont {Agarwal}},\
  }\bibfield  {title} {\enquote {\bibinfo {title} {Collective multiphoton
  blockade in cavity quantum electrodynamics},}\ }\href {\doibase
  10.1103/PhysRevA.95.063842} {\bibfield  {journal} {\bibinfo  {journal} {Phys.
  Rev. A}\ }\textbf {\bibinfo {volume} {95}},\ \bibinfo {pages} {063842}
  (\bibinfo {year} {2017})}\BibitemShut {NoStop}%
\bibitem [{\citenamefont {Lin}\ \emph {et~al.}(2019)\citenamefont {Lin},
  \citenamefont {Hou}, \citenamefont {Zhu},\ and\ \citenamefont
  {Yang}}]{PhysRevA.99.053850}%
  \BibitemOpen
  \bibfield  {author} {\bibinfo {author} {\bibfnamefont {J.~Z.}\ \bibnamefont
  {Lin}}, \bibinfo {author} {\bibfnamefont {K.}~\bibnamefont {Hou}}, \bibinfo
  {author} {\bibfnamefont {C.~J.}\ \bibnamefont {Zhu}}, \ and\ \bibinfo
  {author} {\bibfnamefont {Y.~P.}\ \bibnamefont {Yang}},\ }\bibfield  {title}
  {\enquote {\bibinfo {title} {Manipulation and improvement of multiphoton
  blockade in a cavity-qed system with two cascade three-level atoms},}\ }\href
  {\doibase 10.1103/PhysRevA.99.053850} {\bibfield  {journal} {\bibinfo
  {journal} {Phys. Rev. A}\ }\textbf {\bibinfo {volume} {99}},\ \bibinfo
  {pages} {053850} (\bibinfo {year} {2019})}\BibitemShut {NoStop}%
\bibitem [{\citenamefont {Rabl}(2011)}]{PhysRevLett.107.063601}%
  \BibitemOpen
  \bibfield  {author} {\bibinfo {author} {\bibfnamefont {P.}~\bibnamefont
  {Rabl}},\ }\bibfield  {title} {\enquote {\bibinfo {title} {Photon blockade
  effect in optomechanical systems},}\ }\href {\doibase
  10.1103/PhysRevLett.107.063601} {\bibfield  {journal} {\bibinfo  {journal}
  {Phys. Rev. Lett.}\ }\textbf {\bibinfo {volume} {107}},\ \bibinfo {pages}
  {063601} (\bibinfo {year} {2011})}\BibitemShut {NoStop}%
\bibitem [{\citenamefont {Liao}\ and\ \citenamefont
  {Nori}(2013)}]{PhysRevA.88.023853}%
  \BibitemOpen
  \bibfield  {author} {\bibinfo {author} {\bibfnamefont {Jie-Qiao}\
  \bibnamefont {Liao}}\ and\ \bibinfo {author} {\bibfnamefont {Franco}\
  \bibnamefont {Nori}},\ }\bibfield  {title} {\enquote {\bibinfo {title}
  {Photon blockade in quadratically coupled optomechanical systems},}\ }\href
  {\doibase 10.1103/PhysRevA.88.023853} {\bibfield  {journal} {\bibinfo
  {journal} {Phys. Rev. A}\ }\textbf {\bibinfo {volume} {88}},\ \bibinfo
  {pages} {023853} (\bibinfo {year} {2013})}\BibitemShut {NoStop}%
\bibitem [{\citenamefont {Xie}\ \emph {et~al.}(2017)\citenamefont {Xie},
  \citenamefont {Liao}, \citenamefont {Shang}, \citenamefont {Ye},\ and\
  \citenamefont {Lin}}]{PhysRevA.96.013861}%
  \BibitemOpen
  \bibfield  {author} {\bibinfo {author} {\bibfnamefont {Hong}\ \bibnamefont
  {Xie}}, \bibinfo {author} {\bibfnamefont {Chang-Geng}\ \bibnamefont {Liao}},
  \bibinfo {author} {\bibfnamefont {Xiao}\ \bibnamefont {Shang}}, \bibinfo
  {author} {\bibfnamefont {Ming-Yong}\ \bibnamefont {Ye}}, \ and\ \bibinfo
  {author} {\bibfnamefont {Xiu-Min}\ \bibnamefont {Lin}},\ }\bibfield  {title}
  {\enquote {\bibinfo {title} {Phonon blockade in a quadratically coupled
  optomechanical system},}\ }\href {\doibase 10.1103/PhysRevA.96.013861}
  {\bibfield  {journal} {\bibinfo  {journal} {Phys. Rev. A}\ }\textbf {\bibinfo
  {volume} {96}},\ \bibinfo {pages} {013861} (\bibinfo {year}
  {2017})}\BibitemShut {NoStop}%
\bibitem [{\citenamefont {Hoffman}\ \emph {et~al.}(2011)\citenamefont
  {Hoffman}, \citenamefont {Srinivasan}, \citenamefont {Schmidt}, \citenamefont
  {Spietz}, \citenamefont {Aumentado}, \citenamefont {T\"ureci},\ and\
  \citenamefont {Houck}}]{PhysRevLett.107.053602}%
  \BibitemOpen
  \bibfield  {author} {\bibinfo {author} {\bibfnamefont {A.~J.}\ \bibnamefont
  {Hoffman}}, \bibinfo {author} {\bibfnamefont {S.~J.}\ \bibnamefont
  {Srinivasan}}, \bibinfo {author} {\bibfnamefont {S.}~\bibnamefont {Schmidt}},
  \bibinfo {author} {\bibfnamefont {L.}~\bibnamefont {Spietz}}, \bibinfo
  {author} {\bibfnamefont {J.}~\bibnamefont {Aumentado}}, \bibinfo {author}
  {\bibfnamefont {H.~E.}\ \bibnamefont {T\"ureci}}, \ and\ \bibinfo {author}
  {\bibfnamefont {A.~A.}\ \bibnamefont {Houck}},\ }\bibfield  {title} {\enquote
  {\bibinfo {title} {Dispersive photon blockade in a superconducting
  circuit},}\ }\href {\doibase 10.1103/PhysRevLett.107.053602} {\bibfield
  {journal} {\bibinfo  {journal} {Phys. Rev. Lett.}\ }\textbf {\bibinfo
  {volume} {107}},\ \bibinfo {pages} {053602} (\bibinfo {year}
  {2011})}\BibitemShut {NoStop}%
\bibitem [{\citenamefont {Lang}\ \emph {et~al.}(2011)\citenamefont {Lang},
  \citenamefont {Bozyigit}, \citenamefont {Eichler}, \citenamefont {Steffen},
  \citenamefont {Fink}, \citenamefont {Abdumalikov}, \citenamefont {Baur},
  \citenamefont {Filipp}, \citenamefont {da~Silva}, \citenamefont {Blais},\
  and\ \citenamefont {Wallraff}}]{PhysRevLett.106.243601}%
  \BibitemOpen
  \bibfield  {author} {\bibinfo {author} {\bibfnamefont {C.}~\bibnamefont
  {Lang}}, \bibinfo {author} {\bibfnamefont {D.}~\bibnamefont {Bozyigit}},
  \bibinfo {author} {\bibfnamefont {C.}~\bibnamefont {Eichler}}, \bibinfo
  {author} {\bibfnamefont {L.}~\bibnamefont {Steffen}}, \bibinfo {author}
  {\bibfnamefont {J.~M.}\ \bibnamefont {Fink}}, \bibinfo {author}
  {\bibfnamefont {A.~A.}\ \bibnamefont {Abdumalikov}}, \bibinfo {author}
  {\bibfnamefont {M.}~\bibnamefont {Baur}}, \bibinfo {author} {\bibfnamefont
  {S.}~\bibnamefont {Filipp}}, \bibinfo {author} {\bibfnamefont {M.~P.}\
  \bibnamefont {da~Silva}}, \bibinfo {author} {\bibfnamefont {A.}~\bibnamefont
  {Blais}}, \ and\ \bibinfo {author} {\bibfnamefont {A.}~\bibnamefont
  {Wallraff}},\ }\bibfield  {title} {\enquote {\bibinfo {title} {Observation of
  resonant photon blockade at microwave frequencies using correlation function
  measurements},}\ }\href {\doibase 10.1103/PhysRevLett.106.243601} {\bibfield
  {journal} {\bibinfo  {journal} {Phys. Rev. Lett.}\ }\textbf {\bibinfo
  {volume} {106}},\ \bibinfo {pages} {243601} (\bibinfo {year}
  {2011})}\BibitemShut {NoStop}%
\bibitem [{\citenamefont {Liu}\ \emph {et~al.}(2014)\citenamefont {Liu},
  \citenamefont {Xu}, \citenamefont {Miranowicz},\ and\ \citenamefont
  {Nori}}]{PhysRevA.89.043818}%
  \BibitemOpen
  \bibfield  {author} {\bibinfo {author} {\bibfnamefont {Yu-xi}\ \bibnamefont
  {Liu}}, \bibinfo {author} {\bibfnamefont {Xun-Wei}\ \bibnamefont {Xu}},
  \bibinfo {author} {\bibfnamefont {Adam}\ \bibnamefont {Miranowicz}}, \ and\
  \bibinfo {author} {\bibfnamefont {Franco}\ \bibnamefont {Nori}},\ }\bibfield
  {title} {\enquote {\bibinfo {title} {From blockade to transparency:
  Controllable photon transmission through a circuit-qed system},}\ }\href
  {\doibase 10.1103/PhysRevA.89.043818} {\bibfield  {journal} {\bibinfo
  {journal} {Phys. Rev. A}\ }\textbf {\bibinfo {volume} {89}},\ \bibinfo
  {pages} {043818} (\bibinfo {year} {2014})}\BibitemShut {NoStop}%
\bibitem [{\citenamefont {Miranowicz}\ \emph {et~al.}(2013)\citenamefont
  {Miranowicz}, \citenamefont {Paprzycka}, \citenamefont {Liu}, \citenamefont
  {Bajer},\ and\ \citenamefont {Nori}}]{PhysRevA.87.023809}%
  \BibitemOpen
  \bibfield  {author} {\bibinfo {author} {\bibfnamefont {Adam}\ \bibnamefont
  {Miranowicz}}, \bibinfo {author} {\bibfnamefont {Ma\l{}gorzata}\ \bibnamefont
  {Paprzycka}}, \bibinfo {author} {\bibfnamefont {Yu-xi}\ \bibnamefont {Liu}},
  \bibinfo {author} {\bibfnamefont {Ji\ifmmode
  \check{r}\else~\v{r}\fi{}\'{\i}}\ \bibnamefont {Bajer}}, \ and\ \bibinfo
  {author} {\bibfnamefont {Franco}\ \bibnamefont {Nori}},\ }\bibfield  {title}
  {\enquote {\bibinfo {title} {Two-photon and three-photon blockades in driven
  nonlinear systems},}\ }\href {\doibase 10.1103/PhysRevA.87.023809} {\bibfield
   {journal} {\bibinfo  {journal} {Phys. Rev. A}\ }\textbf {\bibinfo {volume}
  {87}},\ \bibinfo {pages} {023809} (\bibinfo {year} {2013})}\BibitemShut
  {NoStop}%
\bibitem [{\citenamefont {Hovsepyan}\ \emph {et~al.}(2014)\citenamefont
  {Hovsepyan}, \citenamefont {Shahinyan},\ and\ \citenamefont
  {Kryuchkyan}}]{PhysRevA.90.013839}%
  \BibitemOpen
  \bibfield  {author} {\bibinfo {author} {\bibfnamefont {GH}~\bibnamefont
  {Hovsepyan}}, \bibinfo {author} {\bibfnamefont {AR}~\bibnamefont
  {Shahinyan}}, \ and\ \bibinfo {author} {\bibfnamefont {G~Yu}\ \bibnamefont
  {Kryuchkyan}},\ }\bibfield  {title} {\enquote {\bibinfo {title} {Multiphoton
  blockades in pulsed regimes beyond stationary limits},}\ }\href
  {https://journals.aps.org/pra/abstract/10.1103/PhysRevA.90.013839} {\bibfield
   {journal} {\bibinfo  {journal} {Phys. Rev. A}\ }\textbf {\bibinfo {volume}
  {90}},\ \bibinfo {pages} {013839} (\bibinfo {year} {2014})}\BibitemShut
  {NoStop}%
\bibitem [{\citenamefont {Huang}\ \emph {et~al.}(2018)\citenamefont {Huang},
  \citenamefont {Miranowicz}, \citenamefont {Liao}, \citenamefont {Nori},\ and\
  \citenamefont {Jing}}]{PhysRevLett.121.153601}%
  \BibitemOpen
  \bibfield  {author} {\bibinfo {author} {\bibfnamefont {Ran}\ \bibnamefont
  {Huang}}, \bibinfo {author} {\bibfnamefont {Adam}\ \bibnamefont
  {Miranowicz}}, \bibinfo {author} {\bibfnamefont {Jie-Qiao}\ \bibnamefont
  {Liao}}, \bibinfo {author} {\bibfnamefont {Franco}\ \bibnamefont {Nori}}, \
  and\ \bibinfo {author} {\bibfnamefont {Hui}\ \bibnamefont {Jing}},\
  }\bibfield  {title} {\enquote {\bibinfo {title} {Nonreciprocal photon
  blockade},}\ }\href
  {https://journals.aps.org/prl/abstract/10.1103/PhysRevLett.121.153601}
  {\bibfield  {journal} {\bibinfo  {journal} {Phys. Rev. Lett.}\ }\textbf
  {\bibinfo {volume} {121}},\ \bibinfo {pages} {153601} (\bibinfo {year}
  {2018})}\BibitemShut {NoStop}%
\bibitem [{\citenamefont {Deng}\ \emph {et~al.}(2006)\citenamefont {Deng},
  \citenamefont {Payne},\ and\ \citenamefont {Garrett}}]{Deng2006}%
  \BibitemOpen
  \bibfield  {author} {\bibinfo {author} {\bibfnamefont {L.}~\bibnamefont
  {Deng}}, \bibinfo {author} {\bibfnamefont {M.G.}\ \bibnamefont {Payne}}, \
  and\ \bibinfo {author} {\bibfnamefont {W.R.}\ \bibnamefont {Garrett}},\
  }\bibfield  {title} {\enquote {\bibinfo {title} {Effects of multi-photon
  interferences from internally generated fields in strongly resonant
  systems},}\ }\href {\doibase https://doi.org/10.1016/j.physrep.2006.03.005}
  {\bibfield  {journal} {\bibinfo  {journal} {Physics Reports}\ }\textbf
  {\bibinfo {volume} {429}},\ \bibinfo {pages} {123} (\bibinfo {year}
  {2006})}\BibitemShut {NoStop}%
\bibitem [{\citenamefont {Liew}\ and\ \citenamefont {Savona}(2010)}]{Liew2011}%
  \BibitemOpen
  \bibfield  {author} {\bibinfo {author} {\bibfnamefont {T.C.H.}\ \bibnamefont
  {Liew}}\ and\ \bibinfo {author} {\bibfnamefont {V.}~\bibnamefont {Savona}},\
  }\bibfield  {title} {\enquote {\bibinfo {title} {Single photons from coupled
  quantum modes},}\ }\href {\doibase 10.1103/PhysRevLett.104.183601} {\bibfield
   {journal} {\bibinfo  {journal} {Phys. Rev. Lett.}\ }\textbf {\bibinfo
  {volume} {104}},\ \bibinfo {pages} {183601} (\bibinfo {year}
  {2010})}\BibitemShut {NoStop}%
\bibitem [{\citenamefont {Flayac}\ and\ \citenamefont
  {Savona}(2017)}]{Flayac2017}%
  \BibitemOpen
  \bibfield  {author} {\bibinfo {author} {\bibfnamefont {H.}~\bibnamefont
  {Flayac}}\ and\ \bibinfo {author} {\bibfnamefont {V.}~\bibnamefont
  {Savona}},\ }\bibfield  {title} {\enquote {\bibinfo {title} {Unconventional
  photon blockade},}\ }\href {\doibase 10.1103/PhysRevA.96.053810} {\bibfield
  {journal} {\bibinfo  {journal} {Phys. Rev. A}\ }\textbf {\bibinfo {volume}
  {96}},\ \bibinfo {pages} {053810} (\bibinfo {year} {2017})}\BibitemShut
  {NoStop}%
\bibitem [{\citenamefont {Bamba}\ \emph {et~al.}(2011)\citenamefont {Bamba},
  \citenamefont {Imamo\ifmmode~\breve{g}\else \u{g}\fi{}lu}, \citenamefont
  {Carusotto},\ and\ \citenamefont {Ciuti}}]{Bamba2011}%
  \BibitemOpen
  \bibfield  {author} {\bibinfo {author} {\bibfnamefont {Motoaki}\ \bibnamefont
  {Bamba}}, \bibinfo {author} {\bibfnamefont {Atac}\ \bibnamefont
  {Imamo\ifmmode~\breve{g}\else \u{g}\fi{}lu}}, \bibinfo {author}
  {\bibfnamefont {Iacopo}\ \bibnamefont {Carusotto}}, \ and\ \bibinfo {author}
  {\bibfnamefont {Cristiano}\ \bibnamefont {Ciuti}},\ }\bibfield  {title}
  {\enquote {\bibinfo {title} {Origin of strong photon antibunching in weakly
  nonlinear photonic molecules},}\ }\href {\doibase 10.1103/PhysRevA.83.021802}
  {\bibfield  {journal} {\bibinfo  {journal} {Phys. Rev. A}\ }\textbf {\bibinfo
  {volume} {83}},\ \bibinfo {pages} {021802} (\bibinfo {year}
  {2011})}\BibitemShut {NoStop}%
\bibitem [{\citenamefont {Snijders}\ \emph {et~al.}(2016)\citenamefont
  {Snijders}, \citenamefont {Frey}, \citenamefont {Norman}, \citenamefont
  {Bakker}, \citenamefont {Langman}, \citenamefont {Gossard}, \citenamefont
  {Bowers}, \citenamefont {Van~Exter}, \citenamefont {Bouwmeester},\ and\
  \citenamefont {L{\"o}ffler}}]{snijders2016purification}%
  \BibitemOpen
  \bibfield  {author} {\bibinfo {author} {\bibfnamefont {H}~\bibnamefont
  {Snijders}}, \bibinfo {author} {\bibfnamefont {JA}~\bibnamefont {Frey}},
  \bibinfo {author} {\bibfnamefont {J}~\bibnamefont {Norman}}, \bibinfo
  {author} {\bibfnamefont {MP}~\bibnamefont {Bakker}}, \bibinfo {author}
  {\bibfnamefont {EC}~\bibnamefont {Langman}}, \bibinfo {author} {\bibfnamefont
  {A}~\bibnamefont {Gossard}}, \bibinfo {author} {\bibfnamefont
  {JE}~\bibnamefont {Bowers}}, \bibinfo {author} {\bibfnamefont
  {MP}~\bibnamefont {Van~Exter}}, \bibinfo {author} {\bibfnamefont
  {D}~\bibnamefont {Bouwmeester}}, \ and\ \bibinfo {author} {\bibfnamefont
  {W}~\bibnamefont {L{\"o}ffler}},\ }\bibfield  {title} {\enquote {\bibinfo
  {title} {Purification of a single-photon nonlinearity},}\ }\href@noop {}
  {\bibfield  {journal} {\bibinfo  {journal} {Nat. Commun.}\ }\textbf {\bibinfo
  {volume} {7}},\ \bibinfo {pages} {12578} (\bibinfo {year}
  {2016})}\BibitemShut {NoStop}%
\bibitem [{\citenamefont {Tang}\ \emph {et~al.}(2019)\citenamefont {Tang},
  \citenamefont {Deng},\ and\ \citenamefont {Lee}}]{tang2019strong}%
  \BibitemOpen
  \bibfield  {author} {\bibinfo {author} {\bibfnamefont {Jing}\ \bibnamefont
  {Tang}}, \bibinfo {author} {\bibfnamefont {Yuangang}\ \bibnamefont {Deng}}, \
  and\ \bibinfo {author} {\bibfnamefont {Chaohong}\ \bibnamefont {Lee}},\
  }\bibfield  {title} {\enquote {\bibinfo {title} {Strong photon blockade
  mediated by optical stark shift in a single-atom--cavity system},}\
  }\href@noop {} {\bibfield  {journal} {\bibinfo  {journal} {Physical Review
  Applied}\ }\textbf {\bibinfo {volume} {12}},\ \bibinfo {pages} {044065}
  (\bibinfo {year} {2019})}\BibitemShut {NoStop}%
\bibitem [{\citenamefont {Tang}\ \emph {et~al.}(2015)\citenamefont {Tang},
  \citenamefont {Geng},\ and\ \citenamefont {Xu}}]{tang2015quantum}%
  \BibitemOpen
  \bibfield  {author} {\bibinfo {author} {\bibfnamefont {Jing}\ \bibnamefont
  {Tang}}, \bibinfo {author} {\bibfnamefont {Weidong}\ \bibnamefont {Geng}}, \
  and\ \bibinfo {author} {\bibfnamefont {Xiulai}\ \bibnamefont {Xu}},\
  }\bibfield  {title} {\enquote {\bibinfo {title} {Quantum interference induced
  photon blockade in a coupled single quantum dot-cavity system},}\ }\href@noop
  {} {\bibfield  {journal} {\bibinfo  {journal} {Sci. Rep.}\ }\textbf {\bibinfo
  {volume} {5}},\ \bibinfo {pages} {9252} (\bibinfo {year} {2015})}\BibitemShut
  {NoStop}%
\bibitem [{\citenamefont {Ferretti}\ \emph {et~al.}(2013)\citenamefont
  {Ferretti}, \citenamefont {Savona},\ and\ \citenamefont
  {Gerace}}]{ferretti2013optimal}%
  \BibitemOpen
  \bibfield  {author} {\bibinfo {author} {\bibfnamefont {Sara}\ \bibnamefont
  {Ferretti}}, \bibinfo {author} {\bibfnamefont {Vincenzo}\ \bibnamefont
  {Savona}}, \ and\ \bibinfo {author} {\bibfnamefont {Dario}\ \bibnamefont
  {Gerace}},\ }\bibfield  {title} {\enquote {\bibinfo {title} {Optimal
  antibunching in passive photonic devices based on coupled nonlinear
  resonators},}\ }\href@noop {} {\bibfield  {journal} {\bibinfo  {journal} {New
  Journal of Physics}\ }\textbf {\bibinfo {volume} {15}},\ \bibinfo {pages}
  {025012} (\bibinfo {year} {2013})}\BibitemShut {NoStop}%
\bibitem [{\citenamefont {Shen}\ \emph {et~al.}(2015)\citenamefont {Shen},
  \citenamefont {Zhou},\ and\ \citenamefont {Yi}}]{Shen2015}%
  \BibitemOpen
  \bibfield  {author} {\bibinfo {author} {\bibfnamefont {H.~Z.}\ \bibnamefont
  {Shen}}, \bibinfo {author} {\bibfnamefont {Y.~H.}\ \bibnamefont {Zhou}}, \
  and\ \bibinfo {author} {\bibfnamefont {X.~X.}\ \bibnamefont {Yi}},\
  }\bibfield  {title} {\enquote {\bibinfo {title} {Tunable photon blockade in
  coupled semiconductor cavities},}\ }\href {\doibase
  10.1103/PhysRevA.91.063808} {\bibfield  {journal} {\bibinfo  {journal} {Phys.
  Rev. A}\ }\textbf {\bibinfo {volume} {91}},\ \bibinfo {pages} {063808}
  (\bibinfo {year} {2015})}\BibitemShut {NoStop}%
\bibitem [{\citenamefont {Sarma}\ and\ \citenamefont
  {Sarma}(2017)}]{sarma2017quantum}%
  \BibitemOpen
  \bibfield  {author} {\bibinfo {author} {\bibfnamefont {Bijita}\ \bibnamefont
  {Sarma}}\ and\ \bibinfo {author} {\bibfnamefont {Amarendra~K}\ \bibnamefont
  {Sarma}},\ }\bibfield  {title} {\enquote {\bibinfo {title}
  {Quantum-interference-assisted photon blockade in a cavity via parametric
  interactions},}\ }\href@noop {} {\bibfield  {journal} {\bibinfo  {journal}
  {Phys. Rev. A}\ }\textbf {\bibinfo {volume} {96}},\ \bibinfo {pages} {053827}
  (\bibinfo {year} {2017})}\BibitemShut {NoStop}%
\bibitem [{\citenamefont {Wang}\ \emph {et~al.}(2015)\citenamefont {Wang},
  \citenamefont {Gu}, \citenamefont {Liu}, \citenamefont {Miranowicz},\ and\
  \citenamefont {Nori}}]{wang2015tunable}%
  \BibitemOpen
  \bibfield  {author} {\bibinfo {author} {\bibfnamefont {Hui}\ \bibnamefont
  {Wang}}, \bibinfo {author} {\bibfnamefont {Xiu}\ \bibnamefont {Gu}}, \bibinfo
  {author} {\bibfnamefont {Yu-xi}\ \bibnamefont {Liu}}, \bibinfo {author}
  {\bibfnamefont {Adam}\ \bibnamefont {Miranowicz}}, \ and\ \bibinfo {author}
  {\bibfnamefont {Franco}\ \bibnamefont {Nori}},\ }\bibfield  {title} {\enquote
  {\bibinfo {title} {Tunable photon blockade in a hybrid system consisting of
  an optomechanical device coupled to a two-level system},}\ }\href@noop {}
  {\bibfield  {journal} {\bibinfo  {journal} {Phys. Rev. A}\ }\textbf {\bibinfo
  {volume} {92}},\ \bibinfo {pages} {033806} (\bibinfo {year}
  {2015})}\BibitemShut {NoStop}%
\bibitem [{\citenamefont {Li}\ \emph {et~al.}(2019)\citenamefont {Li},
  \citenamefont {Huang}, \citenamefont {Xu}, \citenamefont {Miranowicz},\ and\
  \citenamefont {Jing}}]{li2019nonreciprocal}%
  \BibitemOpen
  \bibfield  {author} {\bibinfo {author} {\bibfnamefont {Baijun}\ \bibnamefont
  {Li}}, \bibinfo {author} {\bibfnamefont {Ran}\ \bibnamefont {Huang}},
  \bibinfo {author} {\bibfnamefont {Xunwei}\ \bibnamefont {Xu}}, \bibinfo
  {author} {\bibfnamefont {Adam}\ \bibnamefont {Miranowicz}}, \ and\ \bibinfo
  {author} {\bibfnamefont {Hui}\ \bibnamefont {Jing}},\ }\bibfield  {title}
  {\enquote {\bibinfo {title} {Nonreciprocal unconventional photon blockade in
  a spinning optomechanical system},}\ }\href@noop {} {\bibfield  {journal}
  {\bibinfo  {journal} {Photon. Res.}\ }\textbf {\bibinfo {volume} {7}},\
  \bibinfo {pages} {630--641} (\bibinfo {year} {2019})}\BibitemShut {NoStop}%
\bibitem [{\citenamefont {Xu}\ \emph {et~al.}(2016)\citenamefont {Xu},
  \citenamefont {Chen},\ and\ \citenamefont {Liu}}]{xu2016phonon}%
  \BibitemOpen
  \bibfield  {author} {\bibinfo {author} {\bibfnamefont {Xun-Wei}\ \bibnamefont
  {Xu}}, \bibinfo {author} {\bibfnamefont {Ai-Xi}\ \bibnamefont {Chen}}, \ and\
  \bibinfo {author} {\bibfnamefont {Yu-xi}\ \bibnamefont {Liu}},\ }\bibfield
  {title} {\enquote {\bibinfo {title} {Phonon blockade in a nanomechanical
  resonator resonantly coupled to a qubit},}\ }\href@noop {} {\bibfield
  {journal} {\bibinfo  {journal} {Phys. Rev. A}\ }\textbf {\bibinfo {volume}
  {94}},\ \bibinfo {pages} {063853} (\bibinfo {year} {2016})}\BibitemShut
  {NoStop}%
\bibitem [{\citenamefont {Shen}\ \emph {et~al.}(2018)\citenamefont {Shen},
  \citenamefont {Shang}, \citenamefont {Zhou},\ and\ \citenamefont
  {Yi}}]{PhysRevA.98.023856}%
  \BibitemOpen
  \bibfield  {author} {\bibinfo {author} {\bibfnamefont {HZ}~\bibnamefont
  {Shen}}, \bibinfo {author} {\bibfnamefont {Cheng}\ \bibnamefont {Shang}},
  \bibinfo {author} {\bibfnamefont {YH}~\bibnamefont {Zhou}}, \ and\ \bibinfo
  {author} {\bibfnamefont {XX}~\bibnamefont {Yi}},\ }\bibfield  {title}
  {\enquote {\bibinfo {title} {Unconventional single-photon blockade in
  non-markovian systems},}\ }\href
  {https://journals.aps.org/pra/abstract/10.1103/PhysRevA.98.023856} {\bibfield
   {journal} {\bibinfo  {journal} {Phys. Rev. A}\ }\textbf {\bibinfo {volume}
  {98}},\ \bibinfo {pages} {023856} (\bibinfo {year} {2018})}\BibitemShut
  {NoStop}%
\bibitem [{\citenamefont {Radulaski}\ \emph {et~al.}(2017)\citenamefont
  {Radulaski}, \citenamefont {Fischer}, \citenamefont {Lagoudakis},
  \citenamefont {Zhang},\ and\ \citenamefont
  {Vu{\v{c}}kovi{\'c}}}]{radulaski2017photon}%
  \BibitemOpen
  \bibfield  {author} {\bibinfo {author} {\bibfnamefont {Marina}\ \bibnamefont
  {Radulaski}}, \bibinfo {author} {\bibfnamefont {Kevin~A}\ \bibnamefont
  {Fischer}}, \bibinfo {author} {\bibfnamefont {Konstantinos~G}\ \bibnamefont
  {Lagoudakis}}, \bibinfo {author} {\bibfnamefont {Jingyuan~Linda}\
  \bibnamefont {Zhang}}, \ and\ \bibinfo {author} {\bibfnamefont {Jelena}\
  \bibnamefont {Vu{\v{c}}kovi{\'c}}},\ }\bibfield  {title} {\enquote {\bibinfo
  {title} {Photon blockade in two-emitter-cavity systems},}\ }\href@noop {}
  {\bibfield  {journal} {\bibinfo  {journal} {Phys. Rev. A}\ }\textbf {\bibinfo
  {volume} {96}},\ \bibinfo {pages} {011801} (\bibinfo {year}
  {2017})}\BibitemShut {NoStop}%
\bibitem [{\citenamefont {Majumdar}\ \emph {et~al.}(2012)\citenamefont
  {Majumdar}, \citenamefont {Bajcsy}, \citenamefont {Rundquist},\ and\
  \citenamefont {Vu\ifmmode \check{c}\else
  \v{c}\fi{}kovi\ifmmode~\acute{c}\else \'{c}\fi{}}}]{Majumdar2012}%
  \BibitemOpen
  \bibfield  {author} {\bibinfo {author} {\bibfnamefont {Arka}\ \bibnamefont
  {Majumdar}}, \bibinfo {author} {\bibfnamefont {Michal}\ \bibnamefont
  {Bajcsy}}, \bibinfo {author} {\bibfnamefont {Armand}\ \bibnamefont
  {Rundquist}}, \ and\ \bibinfo {author} {\bibfnamefont {Jelena}\ \bibnamefont
  {Vu\ifmmode \check{c}\else \v{c}\fi{}kovi\ifmmode~\acute{c}\else
  \'{c}\fi{}}},\ }\bibfield  {title} {\enquote {\bibinfo {title} {Loss-enabled
  sub-poissonian light generation in a bimodal nanocavity},}\ }\href {\doibase
  10.1103/PhysRevLett.108.183601} {\bibfield  {journal} {\bibinfo  {journal}
  {Phys. Rev. Lett.}\ }\textbf {\bibinfo {volume} {108}},\ \bibinfo {pages}
  {183601} (\bibinfo {year} {2012})}\BibitemShut {NoStop}%
\bibitem [{\citenamefont {Liang}\ \emph {et~al.}(2018)\citenamefont {Liang},
  \citenamefont {Duan}, \citenamefont {Guo}, \citenamefont {Liu}, \citenamefont
  {Guan},\ and\ \citenamefont {Ren}}]{liang2018photon}%
  \BibitemOpen
  \bibfield  {author} {\bibinfo {author} {\bibfnamefont {Xinyun}\ \bibnamefont
  {Liang}}, \bibinfo {author} {\bibfnamefont {Zhenglu}\ \bibnamefont {Duan}},
  \bibinfo {author} {\bibfnamefont {Qin}\ \bibnamefont {Guo}}, \bibinfo
  {author} {\bibfnamefont {Cunjin}\ \bibnamefont {Liu}}, \bibinfo {author}
  {\bibfnamefont {Shengguo}\ \bibnamefont {Guan}}, \ and\ \bibinfo {author}
  {\bibfnamefont {Yi}~\bibnamefont {Ren}},\ }\bibfield  {title} {\enquote
  {\bibinfo {title} {Photon blockade in an atom--cavity system},}\ }\href@noop
  {} {\bibfield  {journal} {\bibinfo  {journal} {arXiv preprint
  arXiv:1811.06690}\ } (\bibinfo {year} {2018})}\BibitemShut {NoStop}%
\bibitem [{\citenamefont {Vaneph}\ \emph {et~al.}(2018)\citenamefont {Vaneph},
  \citenamefont {Morvan}, \citenamefont {Aiello}, \citenamefont {F\'echant},
  \citenamefont {Aprili}, \citenamefont {Gabelli},\ and\ \citenamefont
  {Est\`eve}}]{Vaneph2018}%
  \BibitemOpen
  \bibfield  {author} {\bibinfo {author} {\bibfnamefont {Cyril}\ \bibnamefont
  {Vaneph}}, \bibinfo {author} {\bibfnamefont {Alexis}\ \bibnamefont {Morvan}},
  \bibinfo {author} {\bibfnamefont {Gianluca}\ \bibnamefont {Aiello}}, \bibinfo
  {author} {\bibfnamefont {Mathieu}\ \bibnamefont {F\'echant}}, \bibinfo
  {author} {\bibfnamefont {Marco}\ \bibnamefont {Aprili}}, \bibinfo {author}
  {\bibfnamefont {Julien}\ \bibnamefont {Gabelli}}, \ and\ \bibinfo {author}
  {\bibfnamefont {J\'er\^ome}\ \bibnamefont {Est\`eve}},\ }\bibfield  {title}
  {\enquote {\bibinfo {title} {Observation of the unconventional photon
  blockade in the microwave domain},}\ }\href {\doibase
  10.1103/PhysRevLett.121.043602} {\bibfield  {journal} {\bibinfo  {journal}
  {Phys. Rev. Lett.}\ }\textbf {\bibinfo {volume} {121}},\ \bibinfo {pages}
  {043602} (\bibinfo {year} {2018})}\BibitemShut {NoStop}%
\bibitem [{\citenamefont {Snijders}\ \emph {et~al.}(2018)\citenamefont
  {Snijders}, \citenamefont {Frey}, \citenamefont {Norman}, \citenamefont
  {Flayac}, \citenamefont {Savona}, \citenamefont {Gossard}, \citenamefont
  {Bowers}, \citenamefont {van Exter}, \citenamefont {Bouwmeester},\ and\
  \citenamefont {L\"offler}}]{Snijders2018}%
  \BibitemOpen
  \bibfield  {author} {\bibinfo {author} {\bibfnamefont {H.~J.}\ \bibnamefont
  {Snijders}}, \bibinfo {author} {\bibfnamefont {J.~A.}\ \bibnamefont {Frey}},
  \bibinfo {author} {\bibfnamefont {J.}~\bibnamefont {Norman}}, \bibinfo
  {author} {\bibfnamefont {H.}~\bibnamefont {Flayac}}, \bibinfo {author}
  {\bibfnamefont {V.}~\bibnamefont {Savona}}, \bibinfo {author} {\bibfnamefont
  {A.~C.}\ \bibnamefont {Gossard}}, \bibinfo {author} {\bibfnamefont {J.~E.}\
  \bibnamefont {Bowers}}, \bibinfo {author} {\bibfnamefont {M.~P.}\
  \bibnamefont {van Exter}}, \bibinfo {author} {\bibfnamefont {D.}~\bibnamefont
  {Bouwmeester}}, \ and\ \bibinfo {author} {\bibfnamefont {W.}~\bibnamefont
  {L\"offler}},\ }\bibfield  {title} {\enquote {\bibinfo {title} {Observation
  of the unconventional photon blockade},}\ }\href {\doibase
  10.1103/PhysRevLett.121.043601} {\bibfield  {journal} {\bibinfo  {journal}
  {Phys. Rev. Lett.}\ }\textbf {\bibinfo {volume} {121}},\ \bibinfo {pages}
  {043601} (\bibinfo {year} {2018})}\BibitemShut {NoStop}%
\bibitem [{\citenamefont {Ficek}\ and\ \citenamefont
  {Tana{\'s}}(2002)}]{ficek2002entangled}%
  \BibitemOpen
  \bibfield  {author} {\bibinfo {author} {\bibfnamefont {Z}~\bibnamefont
  {Ficek}}\ and\ \bibinfo {author} {\bibfnamefont {Ryszard}\ \bibnamefont
  {Tana{\'s}}},\ }\bibfield  {title} {\enquote {\bibinfo {title} {Entangled
  states and collective nonclassical effects in two-atom systems},}\
  }\href@noop {} {\bibfield  {journal} {\bibinfo  {journal} {Physics Reports}\
  }\textbf {\bibinfo {volume} {372}},\ \bibinfo {pages} {369--443} (\bibinfo
  {year} {2002})}\BibitemShut {NoStop}%
\bibitem [{\citenamefont {Almutairi}\ \emph {et~al.}(2011)\citenamefont
  {Almutairi}, \citenamefont {Tana\ifmmode~\acute{s}\else \'{s}\fi{}},\ and\
  \citenamefont {Ficek}}]{PhysRevA.84.013831}%
  \BibitemOpen
  \bibfield  {author} {\bibinfo {author} {\bibfnamefont {Khulud}\ \bibnamefont
  {Almutairi}}, \bibinfo {author} {\bibfnamefont {Ryszard}\ \bibnamefont
  {Tana\ifmmode~\acute{s}\else \'{s}\fi{}}}, \ and\ \bibinfo {author}
  {\bibfnamefont {Zbigniew}\ \bibnamefont {Ficek}},\ }\bibfield  {title}
  {\enquote {\bibinfo {title} {Generating two-photon entangled states in a
  driven two-atom system},}\ }\href {\doibase 10.1103/PhysRevA.84.013831}
  {\bibfield  {journal} {\bibinfo  {journal} {Phys. Rev. A}\ }\textbf {\bibinfo
  {volume} {84}},\ \bibinfo {pages} {013831} (\bibinfo {year}
  {2011})}\BibitemShut {NoStop}%
\bibitem [{\citenamefont {Bayer}\ \emph {et~al.}(2001)\citenamefont {Bayer},
  \citenamefont {Hawrylak}, \citenamefont {Hinzer}, \citenamefont {Fafard},
  \citenamefont {Korkusinski}, \citenamefont {Wasilewski}, \citenamefont
  {Stern},\ and\ \citenamefont {Forchel}}]{bayer2001coupling}%
  \BibitemOpen
  \bibfield  {author} {\bibinfo {author} {\bibfnamefont {M}~\bibnamefont
  {Bayer}}, \bibinfo {author} {\bibfnamefont {Pawel}\ \bibnamefont {Hawrylak}},
  \bibinfo {author} {\bibfnamefont {K}~\bibnamefont {Hinzer}}, \bibinfo
  {author} {\bibfnamefont {S}~\bibnamefont {Fafard}}, \bibinfo {author}
  {\bibfnamefont {Marek}\ \bibnamefont {Korkusinski}}, \bibinfo {author}
  {\bibfnamefont {ZR}~\bibnamefont {Wasilewski}}, \bibinfo {author}
  {\bibfnamefont {O}~\bibnamefont {Stern}}, \ and\ \bibinfo {author}
  {\bibfnamefont {A}~\bibnamefont {Forchel}},\ }\bibfield  {title} {\enquote
  {\bibinfo {title} {Coupling and entangling of quantum states in quantum dot
  molecules},}\ }\href@noop {} {\bibfield  {journal} {\bibinfo  {journal}
  {Science}\ }\textbf {\bibinfo {volume} {291}},\ \bibinfo {pages} {451--453}
  (\bibinfo {year} {2001})}\BibitemShut {NoStop}%
\bibitem [{\citenamefont {Cox}\ \emph {et~al.}(2012)\citenamefont {Cox},
  \citenamefont {Singh}, \citenamefont {Gumbs}, \citenamefont {Anton},\ and\
  \citenamefont {Carreno}}]{cox2012dipole}%
  \BibitemOpen
  \bibfield  {author} {\bibinfo {author} {\bibfnamefont {Joel~D}\ \bibnamefont
  {Cox}}, \bibinfo {author} {\bibfnamefont {Mahi~R}\ \bibnamefont {Singh}},
  \bibinfo {author} {\bibfnamefont {Godfrey}\ \bibnamefont {Gumbs}}, \bibinfo
  {author} {\bibfnamefont {Miguel~A}\ \bibnamefont {Anton}}, \ and\ \bibinfo
  {author} {\bibfnamefont {Fernando}\ \bibnamefont {Carreno}},\ }\bibfield
  {title} {\enquote {\bibinfo {title} {Dipole-dipole interaction between a
  quantum dot and a graphene nanodisk},}\ }\href@noop {} {\bibfield  {journal}
  {\bibinfo  {journal} {Physical Review B}\ }\textbf {\bibinfo {volume} {86}},\
  \bibinfo {pages} {125452} (\bibinfo {year} {2012})}\BibitemShut {NoStop}%
\bibitem [{\citenamefont {Saffman}\ and\ \citenamefont
  {Walker}(2005)}]{saffman2005analysis}%
  \BibitemOpen
  \bibfield  {author} {\bibinfo {author} {\bibfnamefont {M}~\bibnamefont
  {Saffman}}\ and\ \bibinfo {author} {\bibfnamefont {TG}~\bibnamefont
  {Walker}},\ }\bibfield  {title} {\enquote {\bibinfo {title} {Analysis of a
  quantum logic device based on dipole-dipole interactions of optically trapped
  rydberg atoms},}\ }\href@noop {} {\bibfield  {journal} {\bibinfo  {journal}
  {Physical Review A}\ }\textbf {\bibinfo {volume} {72}},\ \bibinfo {pages}
  {022347} (\bibinfo {year} {2005})}\BibitemShut {NoStop}%
\bibitem [{\citenamefont {Quijandr{\'\i}a}\ \emph {et~al.}(2018)\citenamefont
  {Quijandr{\'\i}a}, \citenamefont {Naether}, \citenamefont {{\"O}zdemir},
  \citenamefont {Nori},\ and\ \citenamefont {Zueco}}]{quijandria2018pt}%
  \BibitemOpen
  \bibfield  {author} {\bibinfo {author} {\bibfnamefont {Fernando}\
  \bibnamefont {Quijandr{\'\i}a}}, \bibinfo {author} {\bibfnamefont {Uta}\
  \bibnamefont {Naether}}, \bibinfo {author} {\bibfnamefont {Sahin~K}\
  \bibnamefont {{\"O}zdemir}}, \bibinfo {author} {\bibfnamefont {Franco}\
  \bibnamefont {Nori}}, \ and\ \bibinfo {author} {\bibfnamefont {David}\
  \bibnamefont {Zueco}},\ }\bibfield  {title} {\enquote {\bibinfo {title}
  {Pt-symmetric circuit qed},}\ }\href@noop {} {\bibfield  {journal} {\bibinfo
  {journal} {Physical Review A}\ }\textbf {\bibinfo {volume} {97}},\ \bibinfo
  {pages} {053846} (\bibinfo {year} {2018})}\BibitemShut {NoStop}%
\end{thebibliography}%
\end{document}